\documentstyle[12pt]{article}
\def\fun#1#2{\lower3.6pt\vbox{\baselineskip0pt\lineskip.9pt
  \ialign{$\mathsurround=0pt#1\hfil##\hfil$\crcr#2\crcr\sim\crcr}}}
\newskip\humongous \humongous=0pt plus 1000pt minus 1000pt

\newif\ifdtup

\def\oldreffmt#1{\rlap{[#1]} \hbox to 2\parindent{}}

\def\figfmt#1{\rlap{Figure {#1}} \hbox to 1in{}}

\def\beq{\begin{equation}}
\def\eeq{\end{equation}}

\def\bq{\begin{quote}}
\def\eq{\end{quote}}

\relax

\addtocounter{equation}{-1}
\newcommand{\be}{\begin{equation}}
\newcommand{\ee}{\end{equation}}
\begin{document}

\begin{titlepage}
\rightline{ETH-TH/92-34}
\rightline{August 1992}
\begin{center}
{\large{\bf Lie-algebraic approach to the theory of polynomial solutions.\\
III. Differential equations in two real variables and general outlook}}
\footnote{This work was supported by the Swiss National Science Foundation}
\vskip 0.8truecm
 {\bf A.Turbiner}\footnote
{On leave of absence from: Institute for Theoretical and Experimental Physics,
Moscow 117259, Russia\\E-mail: turbiner@cernvm or turbiner@vxcern.cern.ch}
\\
Theoretische Physik, ETH-Honggerberg, CH-8093 Zurich, Switzerland
\\
\end{center}
\vskip 0.8truecm
\begin{center}
{\large ABSTRACT}
\end{center}
\begin{quote}

 Classification theorems for linear differential equations
in two real variables, possessing eigenfunctions in the form of the polynomials
(the generalized Bochner problem) are given. The main result is
based on the consideration of the eigenvalue problem for a polynomial
elements of the universal enveloping algebras of the algebras
$sl_3({\bf R})$,  $sl_2({\bf R}) \oplus sl_2({\bf R})$
and  $gl_2 ({\bf R})\ \triangleright\!\!\!< {\bf R}^{r+1}\ , r>0$ taken
in the "projectivized" representations (in differential operators
 of the first order in two real variables) possessing an  invariant subspace.
General insight to the problem of a description of linear differential
operators  possessing an invariant sub-space with a basis in polynomials
is presented. Connection to the recently-discovered quasi-exactly-solvable
problems is discussed.
\end{quote}

\vfill

\end{titlepage}
\newpage

Take the eigenvalue problem
\be
 T \varphi (x,y) \  = \ \epsilon \varphi (x,y)
\ee
where $T$ is a linear differential operator of two real variables,$ x,y$
and $\epsilon$ is the spectral parameter.

{\bf Definition.} Let us give the name of the {\it generalized Bochner
problem} to the problem of
 classification of the differential equations (0) possessing several
eigenfunctions in the form of polynomials.

In the paper \cite{t1} a general method  has been formulated  for generating
eigenvalue problems for linear differential
 operators, linear matrix differential operators and
linear finite-difference operators in one and several variables
possessing polynomial solutions. The
method was based on considering the eigenvalue problem for the
representation of a polynomial element
of the universal enveloping algebra of the Lie algebra in a
finite-dimensional, 'projectivized'
representation of this Lie algebra \cite{t1}.

In two previous papers \cite{t2,t3} it has been proven that in this approach
the consideration of algebras $sl_2(\bf R)$, $sl_2(\bf R)_q$, $osp(2,2)$
in projectivized representations (in differential operators of the first
order) provides
in general both necessary and sufficient conditions for the existance
of polynomial eigenfunctions for ordinary linear differential operators,
finite-difference  operators in one variable \cite{t2} and differential
operators in two variables (one real and one Grassmann)
(or, equivalently, 2 x 2 matrix differential operators) \cite{t3},
respectevely. Particularly, it manifested the classification theorems,
which imply the solution of the original Bochner problem (1929) posed
for ordinary differential equations. In the present paper  similar
classfication theorems will be given for finite-order linear differential
operators in two real variables in connection to the algebras
 $sl_3({\bf R})$,  $sl_2({\bf R}) \oplus sl_2({\bf R})$
and  $gl_2 ({\bf R})\ \triangleright\!\!\!< {\bf R}^m $ . Also  an outlook
on the problem for a general linear differential operators will be
given.

For future considerations, we
define the space of all polynomials in $x,y$ of finite degree as
\be
\label{e1}
{\cal P}_{N,M} \ = \ \langle 1,x^1,y^1,x^2,xy,y^2,\dots,x^N y^M \rangle
\ee
where $N, M$ are non-negative integers, $x,y \in {\bf R}$ are real
variables. It is convenient to illustrate the space of polynomials of
finite degree through the Newton diagrams. In order to do this, let us
consider two-dimensional plane and each point with the integer coordinates
$(k,p)$ put in correspondance with the monomial $x^k y^p$.\par

\section{Polynomials in two real variables}
\subsection{Polynomials of the first type }

\indent
Now let us describe the projectivized representation of the
algebra $sl_3({\bf R})$ in the differential operators of the first order
acting on functions of two real variables. It is easy to show that the
generators have the form \cite{st}
\label{e2}
\[ J^1_3\  =\ y^2 \partial_y \  + \ xy\partial_x\  - \ ny \ , \
J^1_2\  =\ x^2 \partial_x \  + \ xy\partial_y\  - \ nx \ ,\]
\[ J^2_3\  =\ -y \partial_x \ ,\ J^2_1\  =\ - \partial_x \ , \
J^3_1\  =\ - \partial_y \ ,\ J^3_2\  =\ -x \partial_y \ , \]
\be
 J_d\  =\ y \partial_y \  + \ 2x\partial_x\  - \ n \ , \
\tilde J_d\  =\ 2y \partial_y \  + \ x\partial_x\  - \ n \ ,
\ee
where $x,y$ are the real variables and $n$ is a real number. If $n$ is a
 non-negative integer, the representation becomes finite-dimensional of the
dimension $(1+n)(1+n/2)$. The invariant sub-space has a polynomial
basis and is presented as a space of all polynomials of the following type
\be
\label{e3}
{\cal P}^{(I)}_n = \langle 1; x, y; x^2, xy,  y^2;\dots ; x^n, x^{n-1}y,
\dots, x y^{n-1} , y^n \rangle
\ee
or, graphically, (3) is given by the  Newton diagram of the figure 1.

\begin{picture}(400,70)(-10,-20)
\linethickness{1.2pt}
\put(150,10){\line(1,1){30}}
\put(150,10){\line(-1,1){30}}
\linethickness{0.8pt}
\put(120,40){\line(1,0){60}}
\put(152,1){$1$}
\put(110,40){$y^n$}
\put(185,40){$x^n$}
\put(120,40){\circle*{5}}
\put(180,40){\circle*{5}}
\put(150,10){\circle*{5}}
\end{picture}
\begin{center}
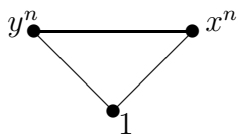
Fig. 1 . \ Graphical representation (Newton diagram) of the space
${\cal P}^{(I)}_n$\linebreak (see (3)).
\end{center}

{\bf Definition.} Let us name a linear differential operator of the $k$-th
order a {\bf quasi-exactly-solvable of the first type},${\bf T}_k^{(I)}(x,y)$,
if it preserves the space ${\cal P}^{(I)}_n$ . Correspondingly, the operator
 ${\bf E}_k^{(I)}(x,y) \in {\bf T}_k^{(I)}(x,y)$, which preserves
the infinite flag $ {\cal P}^{(I)}_0 \subset  {\cal P}^{(I)}_1
 \subset {\cal P}^{(I)}_2
\subset \dots \subset {\cal P}^{(I)}_n \subset \dots$ of spaces of all
polynomials of the type (3) , is named an
{\bf exactly-solvable of the first type}.

{\bf LEMMA 1.1 } {\it Take the space ${\cal P}^{(I)}_n$.

 (i) Suppose $n > (k-1)$.  Any quasi-exactly-solvable operator of the
first type of the $k$-th order ${\bf T}_k^{(I)}(x,y)$, can be
represented by a $k$-th degree polynomial of the generators (2).
 If $n \leq (k-1)$, the part of the quasi-exactly-solvable operator
${\bf T}_k^{(I)}(x,y)$ of the first type containing
derivatives in $x,y$ up to the order $n$ can be represented by a $n$-th
degree polynomial in the generators (2).

(ii) Inversely, any polynomial in (2) is a quasi-exactly solvable operator
of the first type.

(iii) Among quasi-exactly-solvable operators of the first type
there exist exactly-solvable operators of the first type
${\bf E}_k^{(I)}(x,y) \subset {\bf T}_k^{(I)}(x,y)$.}\par

{\it Comment 1.} If we define the universal enveloping algebra $U_g$
of a Lie algebra $g$ as the algebra of all polynomials in generators,
then the meaning of the Lemma is the following: ${\bf T}_k^{(I)}(x,y)$
at $k < n+1$ is simply an element of the universal enveloping algebra
$U_{sl_3({\bf R})}$ of the algebra $sl_3({\bf R})$ in representation (2).
If $k \geq n+1$, then  ${\bf T}_k (x,y)$ is
represented as a polynomial of the $n$-th degree in (2) plus
$B {\partial ^{n+1} \over {\partial x^{n-m+1} \partial y^{m}}}$  , where
$m=0,1, \dots (n+1)$ and $B$ is any linear differential operator
of the order not higher than $(k-n-1)$. \par

{\bf Proof.} The proof is based on the proof of irreducibility (3) and then on
 the application of Burnside theorem \footnote{I appreciate to
J.Frohlich and M.Shubin for suggestion this way of the proof, which is
easy and natural unlike the technical proof originally given at \cite{t2}.
Similarly, one can prove the analogous lemmas for the case of linear
differential operators in one variable, (see Lemma 1 in \cite{t2}),
 finite-difference operators in one variable (Lemma 4 in \cite{t2})
 and differential operators in one real and one
Grassmann variable (Lemma 1 in \cite{t3})} .

Let us introduce the grading of the  generators (2) in the following way. The
 generators are characterized by the two-dimensional grading vectors
\label{e4}
\[ deg (J^1_3) = (+1,0) \ , \ deg (J^1_2) = (0,+1) \ , \]
\[ deg (J^2_3) = (-1,+1) \ , \ deg (J^3_2) = (+1,-1) \ , \]
\[ deg (J_d) = (0,0) \ , \ deg (\tilde J_d) = (0,0) \ ,\]
\be
deg (J^3_1) = (0,-1) \ , \ deg (J^2_1) = (-1,0) .
\ee
Apparently, the grading vector of a monomial in the generators (2) can be
 defined by the grading vectors of the generators by the rule
\label{e5}
\[ \tilde{deg} T \equiv deg [(J^1_3)^{n_{13}} (J^1_2)^{n_{12}}(J^2_3)^{n_{23}}
 (J^3_2)^{n_{32}}
(J_d)^{n_{d}} ({\tilde J}_d)^{n_{\tilde d}}(J^3_1)^{n_{31}} (J^2_1)^{n_{21}}]
 \  = \]
\be
  (n_{13} + n_{32} - n_{23} - n_{31} \ , \ n_{12} + n_{23} - n_{32} - n_{21})
 \equiv (deg_x T\ ,\ deg_y T)
\ee
Here the $n$'s can be arbitrary  non-negative integers.\par
{\bf Definition.} Let us name the grading
of a monomial $T$ in generators (2) the number
\[ deg(T)= deg_x T + deg_y T\ . \]
We will say, that a monomial $T$ possesses positive grading, if this number is
positive. If this number is zero, then a monomial has zero grading.
The notion of grading allows one to classify the operators ${\bf T}_k(x,y)$ in
a
Lie-algebraic sense.\par

{\bf LEMMA 1.2 } {\it A quasi-exactly-solvable operator  ${\bf T}_k^{(I)}
\subset
U_{sl_3({\bf R})}$ either has no terms of positive grading,
 iff it is an exactly-solvable operator.} \par

It is worth noting that
among exactly-solvable operators there exists a certain important class of
degenerate operators, which preserve an infinite flag of spaces of all
homogeneous polynomials
\be
\label{e6}
\tilde {{\cal P}}^{(I)}_n = \langle  x^n, x^{n-1}y, \dots, x y^{n-1} ,
y^n \rangle
\ee
(represented by a horizontal line in Fig.1).

{\bf LEMMA 1.3 } {\it Linear differential operator  ${\bf T}_k(x,y)$ preserves
the infinite flag $\tilde {\cal P}^{(I)}_0 \subset \tilde {\cal P}^{(I)}_1
 \subset\tilde {\cal P}^{(I)}_2
\subset \dots \subset\tilde {\cal P}^{(I)}_n \subset \dots$ of  spaces of
all polynomials of the type (6), iff it is an exactly-solvable operator
having terms of zero grading only. Any operators of such a type can be
represented as a polynomial in the generators $J^2_3, J^3_2, J_d,\tilde J_d$
(see (2)), which form the algebra $so_3 \oplus R$ . If such an operator
contains only terms with zero grading vectors, this operator preserves any
space of polynomials.} \par

{\bf THEOREM 1.1 } {\it Let $n$ be non-negative integer. In general,
the eigenvalue problem for a linear symmetric
 differential operator in two real variables ${\bf T}_k(x,y)$:}
\be
\label{e7}
 {\bf T}_k(x,y) \varphi(x,y) \ = \ \varepsilon \varphi(x,y)
\ee
{\it has $(n+1)(n/2+1)$ eigenfunctions in the form of a polynomial in
variables $x, y$ belonging to the space (3), iff ${\bf T}_k(x,y)$ is a
symmetric quasi-exactly-solvable operator of the first type.
The problem (7) has an infinite sequence of eigenfunctions in the
form of polynomials of the form (3), iff the operator is a symmetric
exactly-solvable operator.} \par

This theorem gives a general classification of differential equations
\be
\label{e8}
\sum_{m=0}^{k}
 \sum_{i=0}^{i=m} a_{i,m-i}^{(m)} (x,y) {{\partial^m \varphi(x,y)} \over
{\partial x^i \partial y^{m-i} }} \ = \ \varepsilon \varphi(x,y)
\ee
having at least one eigenfunction in the form of polynomial  in $x,y$ of the
type (3). In general, the coefficient functions $a_{i,m-i}^{(m)} (x,y)$
have quite cumbersome functional structure and we do not display them here
 (below we will give their explicit form for ${\bf T}_2(x,y)$). They are
polynomials in $x,y$ of the order $(k+m)$ and always contain a general
inhomogeneous polynomial of the order $m$ as a part.
The explicit expressions for those polynomials are obtained by substituting
(2) into a general, the $k$-th order polynomial element of the universal
enveloping algebra $U_{sl_3({\bf R})}$ of the algebra $sl_3({\bf R})$ .
Thus, the coefficients in the polynomials $a_{i,m-i}^{(m)} (x,y)$ can
be expressed through the coefficients of the $k$-th order polynomial element
of the universal enveloping algebra $U_{sl_3({\bf R})}$. The number of free
parameters of the polynomial solutions is defined by the number of parameters
characterizing  a general, $k$-th order polynomial element of the universal
enveloping algebra $U_{sl_3({\bf R})}$. In counting free parameters a
 certain ordering of generators should be fixed  to avoid double
 counting due to commutation relations. Some relations between generators
should be taken into account, specifically for the given representation (2),
like
\label{e9}
\[ J^1_2 J_d\ -\ 2 J^1_2 \tilde J_d\ -\ 3 J^1_3 J^3_1\ = \ n J^1_2 \]

\[ J^1_3 \tilde J_d\ -\ 2 J^1_3 J_d\ -\ 3 J^1_2 J^2_1\ = \ n J^1_3 \]

\[ J^3_2 J_d\ +\  J^3_2 \tilde J_d\ -\ 3 J^1_2 J^3_1\ = \ (n+3) J^3_2 \]

\[ J^2_3 J_d\ +\  J^2_3 \tilde J_d\ -\ 3 J^1_3 J^2_1\ = \ (n+3) J^2_3 \]

\[ 3 (J^1_2 J^2_1\  + \ J^1_3 J^3_1\ +\ J^3_2 J^2_3)\ +
  \ J_d J_d\ +  \tilde J_d \tilde J_d \ -\ J_d \tilde J_d\
= \  3 J_d \  +\ 3 n\ +\ n^2 \]

\[ 2J_d J_d\ +\ 2 \tilde J_d \tilde J_d\ -\ 5 J_d \tilde J_d\ +\
9 J^3_2 J^2_3\ =\ (n+6) J_d\ +\ (n-3) \tilde J_d\ +\ n^2\ +\ 3n \]

\[ J_d J_d\ +\  \tilde J_d \tilde J_d\ +\ 5 (J^1_2 J^2_1\  + \ J^1_3 J^3_1)\
+\ 2 J^3_2 J^2_3\ =\ (n+3) (J_d\ -\ \tilde J_d) \]

\[ J_d J^2_1\ -\ 2 \tilde J_d J^2_1\ -\ 3 J^2_3 J^3_1\ = \ n J^2_1 \]
\be
 J_d J^3_1\ -\ 2 \tilde J_d J^3_1\ -\ 3 J^3_2 J^2_1\ = \ n J^3_1
\ee
between quadratic expressions in generators (and the ideals generated
by them). For the case of exactly-solvable problems, the coefficient functions
$a_{i,m-i}^{(m)} (x,y)$ take the form
\be
\label{e10}
a_{i,m-i}^{(m)} (x,y) \ = \ \sum_{p,q=0}^{p+q \leq m} a_{i,m-i,p,q} x^p y^q
\ee
with arbitrary coefficients.

Now let us proceed to the case of the second-order differential equations.
The second-order polynomial in the generators (2) can be represented as such
\be
\label{e11}
T_2\ =\ c_{\alpha \beta,\gamma \delta} J^{\alpha}_{\beta} J^{\gamma}_{\delta}
\ +\ c_{\alpha\beta} J^{\alpha}_{\beta}\ +\ c \quad ,
\ee
where we imply summation over all repeating indices; $\alpha, \beta,\gamma,
 \delta$ correspond to the indices of operators in (2) and for the Cartan
generators we suppose both indices simulate $d$ or $\tilde d$, all $c$'s
are set to be real numbers. Taking (9) into account, it is easy to show, that
$T_2$ is characterized by 36 free parameters. Substituting (2) into (11),
we obtain the explicit form of the second-order-quasi-exactly-solvable
operator
\[
{\bf T}_2^{(I)}(x,y)\ =\]

\[ [x^2 P_{2,2}^{xx}(x,y)\ +\ x P_{2,1}^{xx}(x,y)\ +\
\tilde{P}_{2,0}^{xx}(x,y)]{\partial^2 \over \partial x^2}\ +\]

\[ [xy P_{2,2}^{xy}(x,y)\ +\ P_{3,1}^{xy}(x,y)\ +\
\tilde{P}_{2,0}^{xy}(x,y)]{\partial^2 \over \partial x \partial y}\ + \]

\[ [y^2 P_{2,2}^{yy}(x,y)\ +\ y P_{2,1}^{yy}(x,y)\ +\
\tilde{P}_{2,0}^{yy}(x,y)]{\partial^2 \over \partial y^2}\ +\]

\[ [x P_{2,2}^{x}(x,y)\ +\ P_{2,1}^{x}(x,y)\ +\
\tilde{P}_{1,0}^{x}(x,y)]{\partial \over \partial x}\ + \]

\[ [y P_{2,2}^{y}(x,y)\ +\ P_{2,1}^{y}(x,y)\ +\
\tilde{P}_{1,0}^{y}(x,y)]{\partial \over \partial y} +\]

\be
\label{e12}
[ P_{2,2}^{0}(x,y)\ +\ P_{1,1}^{0}(x,y)\ +\ \tilde{P}_{0,0}^{0}(x,y)]
\ee

\noindent
where $P_{k,m}^{c}(x,y)$ and $\tilde{P}_{k,m}^{c}(x,y)$ are homogeneous and
inhomogeneous polynomials of the order $k$, respectively, the index
$m$ numerates them, the superscript $\underline c$ characterizes the order
 of derivative,
 which this coefficient function corresponds to. For the case of the
second-order-exactly-solvable operator ${\bf E}_2^{(I)} (x,y)$, the structure
 of coefficient function is similar to (12), except for the fact that all
tildeless polynomials disappear.
 If we denote the number of free parameter of the operator $T$ with the
symbol $par(T)$, then it is easy to show that
\footnote{Recall, that for the case of the
second-order-quasi-exactly-solvable differential operator in one
real variable, the number of free parameters  was equal to 9 (see\cite{t2}),
for the case of one real and one Grassmann variables this number was 25
(see \cite{t3}).}

 \[par({\bf T}_2^{(I)}(x,y))=36 \ . \]
\noindent
while for the case of an exactly-solvable operator
(an infinite sequence of polynomial eigenfunctions in (8))
 \footnote{Recall, that for the case of the second-order-exactly-solvable
differential operator in one real variable, the number of free parameter
was equal to 6 (see \cite{t2}), for the case of one real and one Grassman
variables this number is 17 (see \cite{t3}).}

\[ par({\bf E}_2^{(I)}(x,y))=25 \ . \]

An important particular case is  when the quasi-exactly-solvable operator
${\bf T}_2^{(I)}(x,y)$ possesses two invariant sub-spaces of the type (3).
This situation is described by the following lemma:

{\bf LEMMA 1.4 } {\it Suppose in (11) there are no terms of grading 2:
\be
\label{e13}
c_{12,12}=c_{13,13}=c_{12,13}=0
\ee
and if there exists some coefficients $c$'s and a non-negative integer
 $N$ such that the conditions
\[ c_{12} = (n-N-m)c_{12, d}+(n-2N+m)c_{12,\tilde d}+(N-m)c_{12,32} \ ,\]
\be
\label{e14}
 c_{13} = (n-N-m)c_{13, d}+(n-2N+m)c_{13,\tilde d}+ mc_{13,23} \ ,
\ee
are  fulfilled at all $m=0,1,2,\dots,N$, then the operator
${\bf T}_2^{(I)}(x,y)$ preserves both ${\cal P}_n$ and ${\cal P}_N$, and
$par({\bf T}_2^{(I)}(x,y)) = 31$.} \par

Now let us proceed to the important item: under what conditions on the
coefficients in (11) the second-order-quasi-exactly-solvable operators can be
reduced to a form of the Schroedinger operator after some gauge transformation
\be
\label{e15}
f(x,y)e^{t(x,y)}{\bf T}_2(x,y)e^{-t(x,y)}\ =\ -\Delta_g\ +\ V(x,y)
\ee
where $f, t, V$ are some functions in ${\bf R^2}$, and $\Delta_g$ is the
Laplace-Beltrami operator with some metric tensor $g_{\mu \nu}$;
$ \mu,\nu=1,2$
\footnote{Our further consideration will be restricted the case $f=1$ only}.
 This is a difficult
problem for which there is yet no complete solution. In Refs.\cite{st,gko}
 a few multi-parametric examples were
constructed. However, there is a quite general
situation, for which a rather wide class of the solutions can be obtained.

The algebra $sl_3({\bf R})$ in realization (2) contains the algebra
$so_3({\bf R})$ as a sub-algebra
\[ J^1\  =\ (1+ y^2) \partial_y \  + \ xy\partial_x\  - \ ny \ , \
J^2\  =\ (1+x^2) \partial_x \  + \ xy\partial_y\  - \ nx \ ,\]
\be
\label{e16}
J^3\ =\ x\partial_y \ -\ y\partial_x
\ee
If the parameter $n$ is a non-negative integer (and coincides with
 that in (2)), the same finite-dimensional invariant sub-space
${\cal P}^{(I)}_{n}$ (see (3) and Fig.1) as in the original $sl_3({\bf R})$
occurs. For this case a
corresponding finite-dimensional representation (3) is reducible
and unitary. It has been proven \cite{st,mprst}, that any symmetric
bilinear conbination of generators (16),
$T_2=c_{\alpha \beta}J^{\alpha}J^{\beta},\ c_{\alpha \beta}=c_{\beta \alpha}$,
 can be reduced to a form of the Laplace-Beltrami operator plus a scalar
function. In general, the metric tensor $g_{\mu \nu}$  is not degenerate
\footnote{Degeneracy of $g_{\mu \nu}$ occurs, for example, if
$c_{\alpha \beta}$ has two vanishing
eigenvalues}. Generically, the potential $V(x,y)$ is given by a rational
function and has no dependence on the
spin $n$ (see (16)). However, if the matrix $c_{\alpha \beta}$ has one
vanishing eigenvalue, a certain mysterious relation appears \cite{st,lt}
\be
\label{e17}
V(x,y,\{c\})\ =\ {3 \over 16} R(x,y,\{c\}),
\ee
where $R(x,y,\{c\})$ is the scalar curvature calculated through the metric
tensor attached in the Laplace-Beltrami operator. This can imply that the
corresponding Schroedinger operator has a purely geometrical nature! The real
meaning of this fact is still not understood.

So, we arrive to the two-dimensional exactly-solvable
Schroedinger equations. As a consequence of the quadratic Casimir
operator for $so_3({\bf R})$ in the form (16) being non-trivial and commuting
with $T_2$, the functional space
of $T_2$ is subdivided into the finite-dimensional blocks corresponding to
the irreducible reps of $so_3({\bf R})$.

\subsection{Polynomials of the second type}

In  Ref.\cite{t2} we studied the quasi-exactly-solvable operators in one
real variable. It turned out that the solution to this problem was found using
a connection with the projectivized representation of the algebra
$sl_2({\bf R})$.
As a natural step in developing the original idea, let us consider the
projectivized representation of the direct sum of two
algebras $sl_2({\bf R})$.

The algebra $sl_2({\bf R}) \oplus sl_2({\bf R})$ taken in projectivized
representation acts on functions of two real variables.
The generators have the form (see e.g.\cite{st})
\label{e18}
\[ J_x^+ = x^2 \partial_x - n x\quad ,\quad J_y^+ = y^2 \partial_y - m y \]
\be
 J_x^0 = x \partial_x - {n \over 2} \quad ,\quad
J_y^0 = y \partial_y - {m \over 2}
\ee
\[ J_x^- = \partial_x \quad ,\quad J_y^- = \partial_y \]
where $x,y$ are the real variables and $n,m$ are non-negative integers. There
exists the  finite-dimensional representation of
dimension $(n+1)(m+1)$. Evidently, the invariant sub-space has a polynomial
basis and
is presented as a space of all polynomials with the Newton diagram shown in
Fig.2. We denote this space as ${\cal P}^{(II)}_{n,m}$.

\begin{picture}(400,110)(-10,-20)
\linethickness{1.2pt}
\put(150,10){\line(1,1){50}}
\put(150,10){\line(-1,1){30}}
\linethickness{0.8pt}
\put(120,40){\line(1,1){50}}
\put(200,60){\line(-1,1){30}}
\put(152,1){$1$}
\put(105,40){$y^m$}
\put(165,95){$x^ny^m$}
\put(205,60){$x^n$}
\put(120,40){\circle*{5}}
\put(200,60){\circle*{5}}
\put(170,90){\circle*{5}}
\put(150,10){\circle*{5}}
\end{picture}
\begin{center}
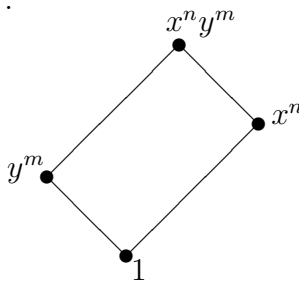
Fig. 2. \ Graphical representation (Newton diagram) of the space
${\cal P}^{(II)}_{n,m}$ .
\end{center}

{\bf Definition.} Let us name a linear differential operator of the $k,p$-th
order, containing derivatives in $x$ and $y$ up to $k-$th
and $p-$th orders, respectively \footnote{So a leading derivative has a form
${\partial^{(k+p)} \over {\partial^k_x\partial^p_y}}$. Also we will use a
notation through this section ${\bf T}_{N}(x,y)$ implying that in general all
derivatives of the order $N$ are presented},
a {\bf quasi-exactly-solvable of the second type}, ${\bf T}_{k,p}^{(II)}(x,y)$,
 if it preserves the space ${\cal P}^{(II)}_{n,m}$. Correspondingly,
the operator ${\bf E}_{k,p}^{(II)}(x,y) \in {\bf T}_{k,p}^{(II)}(x,y)$,
which preserves either
the infinite flag $ {\cal P}^{(II)}_{0,m} \subset  {\cal P}^{(II)}_{1,m}
 \subset {\cal P}^{(II)}_{2,m}
\subset \dots \subset {\cal P}^{(I)}_{n,m} \subset \dots$ , or
the infinite flag $ {\cal P}^{(II)}_{n,0} \subset  {\cal P}^{(II)}_{n,1}
 \subset {\cal P}^{(II)}_{n,2}
\subset \dots \subset {\cal P}^{(I)}_{n,m} \subset \dots$
of  spaces of all polynomials,
is named an {\bf exactly-solvable of the type $2_x$ or $2_y$, respectively}. .

{\bf LEMMA 2.1 } {\it Take the space ${\cal P}^{(II)}_{n,m}$.

 (i) Suppose $n > (k-1)$ and $m > (p-1)$.  Any quasi-exactly-solvable operator
 of the second type $(k,p)$-th order ${\bf T}_{k,p}^{(II)}(x,y)$, can be
represented by a $(k,p)$-th degree polynomial of the generators (18).
 If $n \leq (k-1)$ and/or $m \leq (p-1)$, the part of the
quasi-exactly-solvable operator ${\bf T}_{k,p}^{(II)}(x,y)$ of the second type
containing derivatives in $x,y$ up to the order $n,m$, respectively,
 can be represented by a $(n,m)$-th degree polynomial in the generators (18).

(ii) Inversely, any polynomial in (18) is a  quasi-exactly solvable operator
of the second type.

(iii) Among quasi-exactly-solvable operators of the second type
there exist exactly-solvable operators of the second type ${\bf
E}_{k,p}^{(II)}(x,y)
\subset {\bf T}_{k,p}^{(II)}(x,y)$.}\par

The proof is analogous to the proof of Lemma 1.1 and is based on the
irreducibility of ${\cal P}^{(II)}_{n,m}$ and Burnside theorem.

Similarly, as  has been done for the algebra $sl_3({\bf R})$, one can
introduce the notion of grading:
\label{e19}
\[ deg (J_x^+) = (+1,0) \ , \ deg (J_y^+) = (0,+1) \ , \]
\[ deg (J_x^0) = (0,0) \ , \ deg (J_y^0) = (0,0) \ , \]
\be
deg (J_x^-) = (-1,0) \ , \ deg (J_y^-) = (0,-1) .
\ee
Apparently, the grading vector of a monomial in the generators (18) can be
 defined by the grading vectors (19) of the generators by the rule
\label{e20}
\[ \tilde{deg} T \equiv deg [(J_x^+)^{n_{x+}} (J_x^0)^{n_{x0}}(J_x^-)^{n_{x-}}
(J_y^+)^{n_{y+}} (J_y^0)^{n_{y0}}(J_y^-)^{n_{y-}}] \  = \]
\be
  (n_{x+} - n_{x-} \ , \ n_{y+} - n_{y-}) \equiv (deg_x (T)\ ,\ deg_y (T))
\ee
Here the $n$'s can be arbitrary  non-negative integers.

{\bf Definition.} Let us name the $x-$grading, $y-$grading and grading
of a monomial $T$ in generators (18) the numbers $deg_x (T)$, $deg_y (T)$ and
 $deg(T)= deg_x (T) + deg_y (T)$, respectively.
We say that a monomial $T$ possesses positive $x-$grading ($y-$grading,
 grading), if the number $deg_x (T)$ ($deg_y (T)$, $deg(T)$) is
positive. If $deg_x (T) (deg_y (T), deg(T)) =0$, then a monomial has
zero $x-$grading ($y-$grading, grading).

The notion of grading allows one to classify the operators
${\bf T}_{k,p}^{(II)}(x,y)$ in a Lie-algebraic sense.\par

{\bf LEMMA 2.2 } {\it The quasi-exactly-solvable operator
${\bf T}_{k,p}^{(II)}(x,y)$ of the second type preserves
the infinite flag $ {\cal P}^{(II)}_{0,m} \subset  {\cal P}^{(II)}_{1,m}
 \subset {\cal P}^{(II)}_{2,m}
\subset \dots \subset {\cal P}^{(II)}_{n,m} \subset \dots$
 of spaces of all
the polynomials, iff it is an exactly-solvable operator of the type $2_x$
having no terms of positive $x$-grading, $deg_x > 0$.

The quasi-exactly-solvable
operator  ${\bf T}_{k,p}^{(II)}(x,y)$  preserves
the infinite flag $ {\cal P}^{(II)}_{n,0} \subset  {\cal P}^{(II)}_{n,1}
 \subset {\cal P}^{(II)}_{n,2}
\subset \dots \subset {\cal P}^{(II)}_{n,m} \subset \dots$
 of all spaces of
the polynomials, iff it is an exactly-solvable operator of the type $2_y$
having no terms of positive $y$-grading, $deg_y > 0$.

 If a quasi-exactly-solvable operator of the second type
contains no terms of positive grading, this operator preserves
 the infinite flag $ {\cal P}^{(I)}_0 \subset  {\cal P}^{(I)}_1
 \subset {\cal P}^{(I)}_2
\subset \dots \subset {\cal P}^{(I)}_n \subset \dots$ of spaces of all
the polynomials of the type (3) and is attached to the exactly-solvable
operator of the first type.} \par

{\bf THEOREM 2.1 } {\it Let $n,m$ be non-negative integers. In general,
the eigenvalue problem (7) for a linear symmetric
 differential operator in two real variables ${\bf T}_{k,p}(x,y)$
 has $(n+1)(m+1)$ eigenfunctions in the form of a polynomial in variables
 $x, y$ belonging to the space ${\cal P}^{(II)}_{n,m}$ , iff
${\bf T}_{k,p}(x,y)$
is a quasi-exactly-solvable symmetric operator of the second type.
The problem (7) has an infinite sequence of eigenfunctions in the
form of polynomials belonging the space ${\cal P}^{(II)}_{n,m}$ at fixed $m$
 ($n$), iff the operator is an exactly-solvable symmetric operator of the type
$2_x$ ($2_y$).} \par

This theorem gives a general classification of differential equations (8),
having at least one eigenfunction in the form of a polynomial  in $x,y$ of the
type ${\cal P}^{(II)}_{n,m}$. In general, the coefficient functions
$a_{i,m-i}^{(m)} (x,y)$ in (8)
have a quite cumbersome functional structure and we do not display them here
 (below we will give their explicit form for ${\bf T}_2^{(II)}(x,y)$).
They are polynomials in $x,y$ of the order $(k+m)$.
The explicit expressions for those polynomials are obtained by substituting
(18) into a general, the $k$-th order polynomial element of the universal
enveloping algebra $U_{sl_2({\bf R}) \oplus sl_2({\bf R})}$ of the
algebra $sl_2({\bf R}) \oplus sl_2({\bf R})$ .
Thus, the coefficients in the polynomials $a_{i,m-i}^{(m)} (x,y)$ can
be expressed through the coefficients of the $k$-th order polynomial element
of the universal enveloping algebra $U_{sl_2({\bf R}) \oplus sl_2({\bf R})}$ .
The number of free parameters of the polynomial solutions is defined by the
number of parameters characterizing a general, $k$-th order, polynomial
element of the universal enveloping algebra
$U_{sl_2({\bf R}) \oplus sl_2({\bf R})}$.
In counting free parameters a certain ordering of generators should be fixed
 to avoid double counting due to commutation relations. Also
 some relations between generators should be taken into account, specifically
for the given representation (18), like \cite{t2}
\label{e21}
\[ J^+_x J^-_x\ -\ J^0_x J^0_x\ +\ ({n \over 2} +1) J^0_x\
=\ -{n \over 2}({n \over 2}-1) \]
\be
 J^+_y J^-_y\ -\ J^0_y J^0_y\ +\ ({m \over 2} +1) J^0_y\ =\
-{m \over 2}({m \over 2}-1)
\ee
between quadratic expressions in generators (and the ideals generated by them)
\footnote{For this case they correspond to quadratic Casimir operators}.

Now let us proceed to the case of the second-order differential equations.
The second-order polynomial in the generators (18) can be represented as such
\be
\label{e22}
T_2\ =\ c_{\alpha \beta}^{xx} J^{\alpha}_{x} J^{\beta}_{x}\ +\
c_{\alpha \beta}^{xy} J^{\alpha}_{x} J^{\beta}_{y}\ +\
c_{\alpha \beta}^{yy} J^{\alpha}_{y} J^{\beta}_{y}\ +\
c_{\alpha}^{x} J^{\alpha}_{x}\ +\
 c_{\alpha}^{y} J^{\alpha}_{y}\ +\ c \quad ,
\ee
where we imply summation over all repeating indices and
$\alpha, \beta = \pm,0$; and all $c$'s
are set to be real numbers. Taking (21) in account, it is easy to show that
$T_2$ is characterized by 26 free parameters. Substituting (18) into (22),
we obtain the explicit form of the second-order-quasi-exactly-solvable
operator

\[
{\bf T}_2^{(II)}(x,y)\ =\]

\[  \tilde P_{4,0}^{xx}(x){\partial^2 \over \partial x^2}\ + \
 [x^2y^2 P_{0,2}^{xy}\ +\ xy \tilde P_{1,2}^{xy}(x,y)\ +\
\tilde{P}_{2,0}^{xy}(x,y)]{\partial^2 \over \partial x \partial y}\ + \
 \tilde P_{4,0}^{yy}(y){\partial^2 \over \partial y^2}\ +\]

\be
\label{e23}
 [\tilde P_{3,1}^{x}(x)\  +\
y\tilde{P}_{2,0}^{x}(x)]{\partial \over \partial x}\ + \
 [\tilde P_{3,1}^{y}(y)\  +\
x\tilde{P}_{2,0}^{y}(y)]{\partial \over \partial y}\ + \
 \tilde{P}_{2,0}^{0}(x,y)
\ee

\noindent
where $P_{k,m}^{c}(x,y)$ and $\tilde{P}_{k,m}^{c}(x,y)$ are homogeneous and
inhomogeneous polynomials of the order $k$, respectively, the index
$m$ numerates them, and the superscript $\underline c$ characterizes the order
 of the derivative corresponding to this coefficient function. For the case
of the second-order-exactly-solvable operator of $2_x$ type
\[
{\bf E}_2^{(II)} (x,y) \ = \]
\[  \tilde Q_{2,0}^{xx}(x){\partial^2 \over \partial x^2}\ +\
 [x\tilde{Q}_{2,1}^{xy}(y) \ +\ \tilde{Q}_{2,0}^{xy}(y)]
{\partial^2 \over \partial x \partial y}\ + \
 \tilde Q_{4,0}^{yy}(y){\partial^2 \over \partial y^2}\ +\]
\be
\label{e24}
 \tilde Q_{1,1}^{x}(x)\  +\
y\tilde{Q}_{1,0}^{x}(x)]{\partial \over \partial x}\ + \
 \tilde Q_{1,1}^{y}(y)\  +\
x\tilde{Q}_{1,0}^{y}(y)]{\partial \over \partial y}\ + \
 \tilde{Q}_{2,0}^{0}(y)
\ee
For the case of the second-order-exactly-solvable operator of $2_y$ type
the functional form is similar to (24) with the interchange
$x \leftrightarrow y$.

It is easy to show that the number of free parameters are
\footnote{Recall, that for the case of the
second-order-quasi-exactly-solvable differential operator in one
real variable, the number of free parameters  was equal to 9 (see\cite{t2}),
for the case of one real and one Grassmann variables this number was 25
(see \cite{t3}). Also, for the case of the second-order-exactly-solvable
differential operator in one real variable, the number of free parameter
was equal to 6 (see \cite{t2}), for the case of one real and one Grassmann
variables this number is 17 (see \cite{t3}) (see also p.9 for comparison to
the case of the first type polynomials).}

 \[par({\bf T}_2^{(II)}(x,y))=26 \ . \]
 \[par({\bf E}_2^{(II)}(x,y))=20 \ . \]

Similar to the previous cases, there is a very important particular case of
quasi-exactly-solvable operators of the second order, where they possess two
invariant sub-spaces.

{\bf LEMMA 2.3 } {\it Suppose in (22) there are no terms of $x-$grading 2:
\be
\label{e25}
c_{++}^{xx}=0
\ee
and if there exists some coefficients $c$'s and a non-negative integer
 $N$ such that the conditions
\[ c_{++}^{xy}=0 \ ,\]
\[ c_{+-}^{xy}=0 \ ,\]
\be
\label{e26}
 c_{+}^{x} = (n/2-N)c_{+0}^{xx}+(m/2-k)c_{+0}^{xy} \ ,
\ee
are  fulfilled at all $k=0,1,2,\dots,m$, then the operator
${\bf T}_2^{(II)}(x,y)$
preserves both ${\cal P}_{n,m}^{II}$ and ${\cal P}_{N,m}^{II}$, besides that
$par({\bf T}_2^{(II)}(x,y)) = 22$.} \par

Generically, the question of the reduction of quasi-exactly-solvable operator
of the second type
${\bf T}_2^{(II)}(x,y)$ to the form of the Schroedinger operator is still
open. In the papers \cite{st,gko}  several multi-parametrical families of
those Schroedinger operators were constructed. As in the case of the first type
quasi-exactly-solvable operators, corresponding Schroedinger operators contain
in general the non-trivial Laplace-Beltrami operator.

The above analysis of linear differential operators preserving the space
${\cal P}_{n,m}^{II}$ can be naturally extended to the case of linear
finite-difference operators defined through the Jackson symbol $D$:
\[ D f(x) = {{f(x) - f(qx)} \over {(1 - q) x}} + f(qx) D \]
where $f(x)$ is a real function and $q$ is a number,
instead of continuous derivative. All above-described results hold
\footnote{ with minor modifications} with replacement of the algebra
$sl_2({\bf R}) \oplus sl_2({\bf R})$ to the quantum algebra
$sl_2({\bf R})_q \oplus sl_2({\bf R})_q$ in 'projectivized' representation
(see\cite{t1,t2}). In the limit $q \rightarrow 1$ all results, which can be
obtained, coincide to the results of present Section.

\subsection{Polynomials of the third type}

The third case, which we are going to discuss here, corresponds to the
Lie algebra  $gl_2 ({\bf R})\ \triangleright\!\!\!< {\bf R}^{r+1}$ (semidirect
sum of  $gl_2 ({\bf R})$ with a $(r+1)-$dimensional abelian ideal; the Case 24
in the classification given the paper \cite{gko1}). This family of the
Lie algebras, depending on an integer $r>0$, can be realized in terms of
the first order differential operators
\label{e27}
\[ J^1\  =\  \partial_x \ , \]
\[ J^2\  =\ x \partial_x\ -\ {n \over 3} \ ,\ J^3\  =\ y \partial_y\ -\ {n
\over {3r}} \ , \]
\[ J^4\  =\ x^2 \partial_x \  +\ rxy \partial_y \ - \ nx \ ,\]
\be
 J^{5+i}\  = \ x^{i}\partial_y\ ,\ i=0,1,\dots, r\ ,
\ee
where $x,y$ are the real variables and $n$ is a real number
\footnote{It is worth noting that at $r=1$ the  algebra
$\{ gl_2 ({\bf R})\ \triangleright\!\!\!< {\bf R}^{2}\} \subset sl_3
({\bf R})$. Thus, this case is reduced to one about the first type
polynomials (see Ch.1.1). Hereafter, we include $r=1$ into consideration just
for the purpose of completeness.}.
If $n$ is a non-negative integer, the representation becomes
finite-dimensional. The invariant sub-space has a polynomial basis and
is presented as a space of all polynomials of the form
\be
\label{e28}
{\cal P}^{(III)}_{r,n} = \sum_{i,j \geq 0}^{i+rj \leq n} a_{ij} x^i y^j
\ee
or, graphically, (28) is given by the Newton diagram of the figure 3. The
general formulas for the dimension of corresponding
finite-dimensional representation (28) is given by
\be
\label{e29}
 dim {\cal P}^{(III)}_{r,n}= {{[n^2+(r+2)n+\alpha_{r,n}]} \over {2r}}
\ee
 where for small $r$
\[ \alpha_{1,n}=2 \ , \]
\[ \alpha_{2,n}  = \left\{
\begin{array}{cc}
3 & \mbox{at odd n} \\
4 & \mbox{at even n}
\end{array} \right. \]
\[ \alpha_{3,n}= \left\{
\begin{array}{cc}
4 & \mbox{at (n+1) multiple 3} \\
6 & \mbox{at other n}
\end{array} \right. \]
\[ \alpha_{4,n}= \left\{
\begin{array}{cc}
5 & \mbox{at (n+1) multiple 4} \\
8 & \mbox{at other n}\\
9 & \mbox{at (n+3) multiple 4}
\end{array} \right. \]
Actually, the numbers $\alpha_{r,n}$ are given by all possible products
$m(2r-m)$ at $m=1,2,\dots,r$.

\begin{picture}(400,120)(-10,-20)
\linethickness{1.2pt}
\put(150,10){\line(1,1){60}}
\put(150,10){\line(-1,1){30}}
\linethickness{0.8pt}
\put(120,40){\line(3,1){90}}
\put(152,1){$1$}
\put(108,41){$y^j$}
\put(215,70){$x^i$}
\put(120,40){\circle*{5}}
\put(210,69){\circle*{5}}
\put(150,10){\circle*{5}}
\end{picture}
\begin{center}
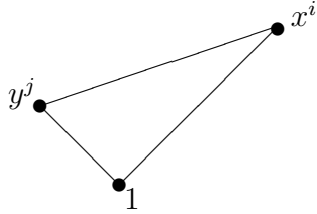
Fig. 3 . \ Graphical representation (Newton diagram) of the space
${\cal P}^{(III)}_{r,n}$\linebreak (see (28)).
\end{center}

{\bf Definition.} Let us name a linear differential operator of the $k$-th
order a {\bf quasi-exactly-solvable of the $r-$third type},
${\bf T}_k^{(r,III)}(x,y)$, if it preserves the space ${\cal P}^{(III)}_{r,n}$
 . Correspondingly, the
operator ${\bf E}_k^{(r,III)}(x,y) \in {\bf T}_k^{(r,III)}(x,y)$, which
preserves
the infinite flag $ {\cal P}^{(III)}_{r,0} \subset  {\cal P}^{(III)}_{r,1}
 \subset {\cal P}^{(III)}_{r,2}
\subset \dots \subset {\cal P}^{(III)}_{r,n} \subset \dots$ of spaces of all
 polynomials of the type (28) , is named an
{\bf exactly-solvable of the $r-$third type}.

{\bf LEMMA 3.1 } {\it Take the space ${\cal P}^{(III)}_{r,n}$.

 (i) Suppose $n > (k-1)$.  Any quasi-exactly-solvable operator of the
$r-$third  type of the $k$-th order ${\bf T}_k^{(r,III)}(x,y)$, can be
represented by a $k$-th degree polynomial of the generators (27).
 If $n \leq (k-1)$, the part of the quasi-exactly-solvable operator
${\bf T}_k^{(r,III)}(x,y)$ of the $r-$third type containing
derivatives in $x,y$ up to the order $n$ can be represented by a $n$-th
degree polynomial in the generators (27).

(ii) Inversely, any polynomial in (27) is a quasi-exactly solvable operator
of the $r-$third type.

(iii) Among quasi-exactly-solvable operators of the $r-$third type
there exist exactly-solvable operators of the $r-$third type
${\bf E}_k^{(r,III)}(x,y) \subset {\bf T}_k^{(r,III)}(x,y)$.}\par

The proof is analogous to the proof of the Lemma 1.1 and 2.1. It is based on
irreducibility ${\cal P}^{(III)}_{r,n}$ and Burnside theorem.

One can introduce the grading of the  generators (27) in analogous way as
has been done before for the cases of the algebra $sl_3({\bf R})$ (see (4))
and $sl_2({\bf R}) \oplus sl_2({\bf R})$ (see (19)). All
 generators are characterized by the two-dimensional grading vectors
\label{e30}
\[ deg (J^1) = (-1,0) \ , \]
\[ deg (J^2) = (0,0) \ , \ deg (J^3) = (0,0) \ , \]
\[ deg (J^4) = (1,0) \ ,\]
\be
deg (J^5) = (0,-1) \ , \ deg (J^6) = (1,-1) \ ,\dots, \ deg (J^{5+r}) =
(r,-1)
\ee
Similarly as before, the grading vector of a monomial in the generators (27)
can be defined through the grading vectors of the generators (27)
(cf.(5),(20)).\par
{\bf Definition.} Let us name the grading
of a monomial $T$ in generators (27) the number
\[ deg(T)= deg_x T + r deg_y T \]
 (cf. the case of $sl_3({\bf R})$).
We will say that a monomial $T$ possesses positive grading if this number is
positive. If this number is zero, then a monomial has zero grading.
The notion of grading allows one to classify the operators ${\bf T}_k(x,y)$ in
a Lie-algebraic sense.\par

{\bf LEMMA 3.2 } {\it A quasi-exactly-solvable operator  ${\bf T}_k^{(r,III)}
 \subset U_{gl_2 ({\bf R})\ \triangleright\!< {\bf R}^{r+1}}$ either
has no terms of positive grading, iff it is an exactly-solvable
operator.} \par

{\bf THEOREM 3.1 } {\it Let $n$ and $(r-1)$ be non-negative integers.
In general, the eigenvalue problem (7) for a linear symmetric
 differential operator of the $k-$th order in two real variables
${\bf T}_k(x,y)$ has a certain amount of eigenfunctions in the form of a
polynomial in variables
 $x, y$ belonging to the space (28), iff ${\bf T}_k(x,y)$ is a
quasi-exactly-solvable, symmetric operator of the $r-$third type.
The problem (7) has an infinite sequence of eigenfunctions in the
form of polynomials belonging to (28), iff the operator is an
exactly-solvable, symmetric operator of the $r-$third type.} \par

This theorem gives a general classification of differential equations (8),
having at least one eigenfunction in the form of a polynomial  in $x,y$ of the
type ${\cal P}^{(III)}_{r,n}$. In general, the coefficient functions
$a_{i,m-i}^{(m)} (x,y)$ in (8) are polynomials in $x,y$ and
have a quite cumbersome functional structure and we do not display them here
 (below we will give their explicit form for ${\bf T}_2^{(r,III)}(x,y)$).
The explicit expressions for those polynomials are obtained by substituting
(27) into a general, $k$-th order polynomial element of the universal
enveloping algebra $U_{gl_2 ({\bf R})\ \triangleright\!< {\bf R}^{r+1}}$
of the algebra $gl_2 ({\bf R})\ \triangleright\!\!\!< {\bf R}^{r+1}$.
Thus, the coefficients in the polynomials $a_{i,m-i}^{(m)} (x,y)$ can
be expressed through the coefficients of the $k$-th order polynomial element
of the universal enveloping algebra
$U_{gl_2 ({\bf R})\ \triangleright\!< {\bf R}^{r+1}}$.
The number of free parameters of the polynomial solutions is defined by the
number of parameters characterizing  general $k$-th order polynomial element
of the universal enveloping algebra.
In counting free parameters a certain ordering of generators should be fixed
to avoid double counting due to commutation relations. Also
 some relations between generators should be taken into account, specifically
for a given representation (27), like
\label{e31}
\[ J^2 J^5\ -\ J^1 J^6\ +\ {n \over 3}J^5\ =\ 0\ , \]
\be
 J^1 J^4\ -\ J^2 J^2\ -\ rJ^2 J^3\ -\ J^2\ -\ r({n \over 3}+1) J^3\ =
\ - {n \over 3}({n \over 3}+1) \ ,
\ee
\label{e32}
\[ J^2 J^{6+i}\ +\ rJ^3 J^{6+i}\ -\ J^4 J^{5+i}\ -\ ({n \over 3} +1)J^{6+i}\ =\
0\ ,\]
\be
 at\ i=0,1,2,\dots,(r-1) \ ,
\ee
\label{e33}
\[ J^1 J^{7+i}\ -\ J^2 J^{6+i}\ -\ ({n \over 3}+1) J^{6+i}\ =\ 0\ ,\]
\be
 at\ i=0,1,2,\dots,(r-2) \ ,
\ee
\label{e34}
\[ J^{5} J^{7+i}\ = \dots =\ J^{5+k} J^{7+i-k}\ , \]
\be
 at\ k=0,1,2,\dots ,\ and\ 2k \leq (2+i)\ and\ i=0,1,2,\dots ,(2r-2)
\ee
between quadratic expressions in generators (and the ideals generated
by them).

Now let us proceed to the case of the second-order differential equations.
The second-order polynomial in the generators (27) can be represented as
\be
\label{e35}
T_2\ =\ c_{\alpha \beta} J^{\alpha} J^{\beta}\ +\
c_{\alpha} J^{\alpha}\ +\ c \quad ,
\ee
where $\alpha, \beta=1,2,\dots,(5+r)$ and we imply summation over
all repeating indices;  all $c$'s
are set to be real numbers. Taking the relations (31)-(34) in account,
one can count the number of free parameters and obtain
\label{e36}
\be
par({\bf T}_2^{(r,III)} (x,y))\ =\ 5(r+4)
\ee
 Substituting (27) into (35),
we obtain the explicit form of the second-order-quasi-exactly-solvable
operator
\[
{\bf T}_2^{(r,III)}(x,y)\ =\]
\[  \tilde P_{4,0}^{xx}(x){\partial^2 \over \partial x^2}\ + \
 [\tilde P_{r+2,1}^{xy}(x)\ +\
y\tilde{P}_{3,0}^{xy}(x)]{\partial^2 \over \partial x \partial y}\ + \
 [\tilde P_{2r,1}^{yy}(x)\ +\
y\tilde{P}_{r+1,0}^{yy}(x)]  ]{\partial^2 \over \partial y^2}\ +\]
\be
\label{e37}
 [\tilde P_{3,0}^{x}(x)\ ]{\partial \over \partial x}\ + \
 [\tilde P_{r+1,1}^{y}(x)\  +\
y\tilde{P}_{1,0}^{x}(x)]{\partial \over \partial y}\ + \
 \tilde{P}_{2,0}^{0}(x)
\ee

\noindent
where  $\tilde{P}_{k,m}^{c}(x,y)$ are inhomogeneous polynomials of the
order $k$, the index
$m$ numerates them, and the superscript $\underline c$ characterizes the order
 of the derivative corresponding to this coefficient function. For the case
of the second-order-exactly-solvable operator of $r$-third type
\[
{\bf E}_2^{(r,III)}(x,y)\ =\]
\[  \tilde P_{2,0}^{xx}(x){\partial^2 \over \partial x^2}\ + \
 [\tilde P_{r+1,1}^{xy}(x)\ +\
y\tilde{P}_{1,0}^{xy}(x)]{\partial^2 \over \partial x \partial y}\ + \
 [\tilde P_{2r,1}^{yy}(x)\ +\
y\tilde{P}_{r,0}^{yy}(x)]  ]{\partial^2 \over \partial y^2}\ +\]
\be
\label{e38}
 [\tilde P_{1,0}^{x}(x)\ ]{\partial \over \partial x}\ + \
 [\tilde P_{r,1}^{y}(x)\  +\
y\tilde{P}_{0,0}^{x}(x)]{\partial \over \partial y}\ + \
 \tilde{P}_{0,0}^{0}(x)
\ee
In this case the number of free parameters is equal to
\label{e39}
\be
par({\bf E}_2^{(r,III)} (x,y))\ =\ 5(r+3)
\ee
(cf.(36)).

Similar to the previous cases, there is an important particular case of
$r$-third type
quasi-exactly-solvable operators of the second order, where they possess two
invariant sub-spaces.

{\bf LEMMA 3.3 } {\it Suppose in (35) there are no terms of grading 2:
\be
\label{e40}
c_{44}=0
\ee
and if there exists some coefficients $c$'s and a non-negative integer
 $N$ such that the conditions
\[ c_{4,5+r}=0 \ ,\]
\be
\label{e41}
 c_{4}\ =\ ({N \over 3}-m) c_{24}\ +\ ({N \over 3}-k)c_{34}\ ,
\ee
are  fulfilled at all $m,k=0,1,2,\dots$ such that $m+rk=N$, then the
operator ${\bf T}_2^{(r,III)}(x,y)$
preserves both ${\cal P}_{r,n}^{III}$ and ${\cal P}_{r,N}^{III}$, and
$par({\bf T}_2^{(r,III)}(x,y)) =5r+17 $.} \par

Generically, the question of the reduction of quasi-exactly-solvable operator
of the $r$-third type
${\bf T}_2^{(r,III)}(x,y)$ to the form of the Schroedinger operator is still
open. Initially \cite{gko} several
multi-parametrical families of those Schroedinger operators were constructed.
As in the case quasi-exactly-solvable operators of the first and the second
type, corresponding Schroedinger operators contain
in general a non-trivial Laplace-Beltrami operator.

\subsection{Discussion}

Through all the above analysis, very crucial requirement played a role: we
considered the problem in general position, assuming the spaces of all
polynomials contain polynomials with all possible coefficients. If this
requirement is not fullfilled,
than the above connection of the finite-dimensional space of all
polynomials and the finite-dimensional representation of the Lie algebra
is lost and degenerate cases occur. This demands a separate investigation.

So, we described three types of finite-dimensional spaces (see Fig.1-3)
with a basis in
polynomials of two real variables, which can be preserved by linear
differential operators. Very natural question emerges: is it possible to find
linear differential operators, which preserve the space of polynomials of
finite order other than shown in Fig.1,2,3 ?

{\bf Conjecture 1.} {\it If a linear differential operator $T_k(x,y)$ does
preserve a finite-dimensional space of all polynomials other than
${\cal P}_{n}^{I}, {\cal P}_{n,m}^{II}, {\cal P}_{r,n}^{III}$, then
this operator preserves also a certain infinite flag of finite-dimensional
spaces of all polynomials.}

\section{General consideration}

The main conclusion of the considerations of the present paper and
the papers \cite{t1,t2,t3} one can formulate as the following theorem:

{\bf (Main) Theorem.} {\it Take a Lie algebra $g$, realized in differential
operators of the first order, which possesses a finite-dimensional irreducible
representation ${\cal P}$ ( a finite module of smooth functions). Any linear
differential operator, having the invariant subspace ${\cal P}$, which
 coincides to the above finite module of smooth functions, can be
represented by a polynomial in generators of the algebra $g$ plus an
operator annuhilating  ${\cal P}$.}

{\bf Proof.}  Use of the Burnside theorem.

As a probable extension of this theorem, we assume that the following
conjecture hold

{\bf Conjecture 2.} {\it If a linear differential operator $T$ acting on
functions in ${\bf R^k}$ does
possess the only finite-dimensional invariant subspace with polynomial basis,
this finite-dimensional space coincides to a certain finite-dimensional
representation of some Lie algebra, realized by differential operators of the
first order.}

The interesting question is how to describe finite modules of smooth
functionsin ${\bf R^k}$, which can serve as invariant sub-spaces of linear
operators.
Evidently, if we could found some realization of a Lie algebra $g$ in
differential operators, possessing an irreducible finite module of smooth
functions, one can immediately consider the direct sum of several species of
$g$ acting on the functions given at the corresponding direct products of
spaces, where the original algebra $g$ acts in each of them. As an example,
this procedure is presented in the Section 1.2 : the direct sum of two
$sl_2({\bf R})$ leads to the operators acting on the functions at ${\bf R^2}$
and a rectangular ${\cal P}^{(II)}_{n,m}$ is preserved. Taking the
direct sum of $k$ species of $sl_2({\bf R})$ acting at ${\bf R^k}$ ,
we arrive to $k$-dimensional parallelopiped as the invariant sub-space.
In this case the algebra $sl_2({\bf R})$ plays the role of a
{\it primary} algebra giving rise to a
multidimensional geometrical figure (Newton diagram) as the invariant sub-space
of some linear differential operators. In the space ${\bf R^1}$ there is only
one such  "irreducible" Newton diagram -- a finite interval, in other words, a
space of all polynomials in $x$ of degree not higher than a certain integer.
Thus, in ${\bf R^2}$ the rectangular Newton diagram is "reducible" stemming
from the product of two intervals as one-dimensional Newton diagrams..
In ${\bf R^2}$ the irreducible Newton diagrams are exhausted by different
triangles, connected to the algebras $sl_3({\bf R})$ and $\{ gl_2 ({\bf R})\
\triangleright\!\!\!< {\bf R}^{r+1}\}$. We could not find any other Newton
diagrams in ${\bf R^2}$ (see conjecture 1).

Before to present our present knowledge about Newton diagrams for the
${\bf R^k}$ case, let us recall on the regular representation
(in the first order differential
operators) of the Lie algebra $sl_N({\bf R})$ given on the flag manifold,
which acts on the smooth functions of $N(N-1)/2$ real variables $z_{i,i+q}\ ,
i=1,2,\dots,(N-1),\ q=1,2,\dots,(N-i)$. The explicit formulas for generators
are given by (see \cite{r})

\label{e}
\[ D(e_i)={\partial \over {\partial z_{i,i+1}}}- \sum_{q=i+2}^{N}
z_{i+1,q}{\partial \over {\partial z_{i,q}}} \]

\[ D(f_i)=\sum_{q=1}^{i}z_{q,i+1}z_{i,i+1}{\partial \over {\partial z_{q,i+1}}}
+
\sum_{q=1}^{i-1}(z_{q,i+1}-z_{q,i}z_{i,i+1}){\partial \over {\partial z_{q,i}}}
\]
\[
\sum_{q=i+2}^{N}z_{i,q}{\partial \over {\partial z_{i+1,q}}} \ - \
n_i z_{i,i+1} \]

\be
 D(\tilde h_i)=-\sum_{q=i+1}^{N} z_{i,q}{\partial \over {\partial z_{i,q}}}+
\sum_{q=1}^{i-1} z_{q,i}{\partial \over {\partial z_{q,i}}}-
\sum_{p=1}^{i} n_{i-p}
\ee
\noindent
where we use a notation $e_i,f_i,\tilde h_i$ for generators of positive and
negative roots, and the Cartan generators, correspondingly. If $n_i$ are
non-negative integers, the finite-dimensional irreducible representation of
$sl_N({\bf R})$ will occur in the form of inhomogeneous polynomials in
variables $z_{i,i+q}$. The highest weight vector is characterized by the
integer numbers $n_i,\ i=1,2,\dots, (N-1)$.

Connected to the space ${\bf R^3}$, at least, two irreducible Newton
diagrams appear :
a tetrahedron, connected to a degenerate finite-dimensional representation
of $sl_4({\bf R})$ (highest weight vector has two vanishing integers and
one non-vanishing), and a certain geometrical figure, corresponding to the
regular representation of $sl_3({\bf R})$ given on the flag manifold. We
do not know if those exhaust all irreducible Newton diagrams or none. For the
general case, ${\bf R^k}$ there also exist polyhedra connected to
$sl_{k+1}({\bf R})$  (except one, all integers, characterizing the highest
weight vector are vanishing) and geometrical figures related to some
finite-dimensional regular representations of $sl_{k+1-i}({\bf R}),\ i>0$.
It is easy to show since the algebra $so_{k+1}({\bf R})
\subset sl_{k+1}({\bf R})$ and has the polyhedron as the invariant sub-space
\footnote{Corresponding finite-dimensional representation is unitary and
reducible (see e.g. the discussion in the end of Section 1.1)}, any
quadratic polynomial in
generators of the algebra $so_{k+1}({\bf R})$ with real symmetric
coefficients can be reduced to the form the Laplace-Beltrami operator plus a
scalar function by means of a gauge transformation (see discussion in Ref.
\cite{mprst}).

The above procedure can be extended to the case of the Lie super-algebras
and also the quantum algebras. For the former, for linear operators acting on
functions of real and Grassmann variables one can appear finite-dimensional
invariant subspaces others than finite-dimensional representations
(see \cite{t3}). However, those subspaces are connected to the
representations of polynomial elements of the universal enveloping algebra
and, correspondingly, they can be constructed through finite-dimensional
representations of the original super-algebra. For the latter, we could not
find any irreducible (non-trivial) Newton diagrams in ${\bf R^k}$ for $k>1$;
they will occur, if a quantum space is considered instead of ordinary
one \footnote{I am grateful to O.V.Ogievetski for discussion of this
topic.}. Also one can construct quasi-exactly-solvable and exactly-solvable
operators considering "mixed" algebras: a direct sum of the Lie algebras, Lie
superalgebras and the quantum algebras, realized in the first order
differential and/or difference operators, possessing a finite-dimensional
 invariant subspace.

We have described above the linear operators, which possess a
finite-dimensional invariant sub-space with a polynomial basis.
Certainly, by means of diffeomorphism and
gauge transformation one can obtain the operators with an invariant subspace
with a non-polynomial basis (but emerging from polynomial one). The open
question is : is it possible to find
the operators possessing a finite-dimensional invariant sub-space with some
explicit basis, which can {\it not} be reduced to a polynomial one using a
diffeomorphism and gauge transformation? Up to now no
examples of this type have been found (see Conjecture 2). Since the
original aim of this whole investigation was the
construction of the quasi-exactly-solvable Schroedinger operators
\cite{t4,st,mprst,olver,gko,lt}, a resolution of this problem plays a crucial
 role in leading to a complete classification of the
quasi-exactly-solvable Schroedinger operators.

\begin{center}
{\bf ACKNOLEDGEMENTS}

\end{center}

At closing, I am very indebted to V.Arnold, J.Frohlich and M.Shubin for
the interest to the subject and useful discussions. Also I am grateful
to the Institute of Theoretical Physics, ETH-Honggerberg, where
this work was completed, for their kind hospitality extended to me.

\end{document}